\documentclass[%
reprint,
superscriptaddress,
amsmath,amssymb,
aps,
]{revtex4-1}
\usepackage{soul}
\usepackage{float}
\usepackage{graphicx}
\usepackage[caption=false]{subfig}
\usepackage{dcolumn}
\usepackage{bm}
\usepackage[inline]{enumitem}
\usepackage{bbm}
\usepackage{xcolor}
\usepackage{soul}

\newcommand{\dd}{{\rm d}}
\newcommand{\brho}{{\tilde \rho}}
\newcommand{\dt}{{\dd t}}
\newcommand{\ddt}{{\delta t}}

\definecolor{nblue}{rgb}{0.06,0.3,0.73}
\definecolor{nblack}{rgb}{0,0,0}
\definecolor{nred}{rgb}{0.9,0.1,0.1}
\definecolor{nmagenta}{rgb}{0.7,0.0,0.3}
\definecolor{neditcolor}{rgb}{0.3,0.3,0.9}

\begin{document}

	\title{Time-delayed quantum feedback and incomplete decoherence suppression\\ with no-knowledge measurement}%
	
	\author{Jirawat Saiphet}
	\email{jirawat.sai@outlook.com}
	\affiliation{Optical and Quantum Physics Laboratory, Department of Physics, Faculty of Science, Mahidol University, Bangkok, 10400, Thailand}%
	\author{Sujin Suwanna}
	\affiliation{Optical and Quantum Physics Laboratory, Department of Physics, Faculty of Science, Mahidol University, Bangkok, 10400, Thailand}%
	\author{Andr\'e R. R. Carvalho}
	\affiliation{Centre for Quantum Dynamics, Griffith University, Nathan, Queensland, 4111, Australia}
	\altaffiliation{Current address: Q-CTRL Sydney, NSW, Australia}
	\author{Areeya Chantasri}
	\email{areeya.chn@mahidol.ac.th}
	\affiliation{Optical and Quantum Physics Laboratory, Department of Physics, Faculty of Science, Mahidol University, Bangkok, 10400, Thailand}
	\affiliation{Centre for Quantum Dynamics, Griffith University, Nathan, Queensland, 4111, Australia}
	
	
	\date{\today}
	
	\begin{abstract}
		The no-knowledge quantum feedback was proposed by Szigeti \emph{et al.}, Phys.~Rev.~Lett.~\textbf{113}, 020407 (2014), as a measurement-based feedback protocol for decoherence suppression for an open quantum system. By continuously measuring environmental noises and feeding back controls on the system, the protocol can completely reverse the measurement backaction and therefore suppress the system's decoherence. However, the complete decoherence cancellation was shown only for the instantaneous feedback, which is impractical in real experiments. Therefore, in this work, we generalize the original work and investigate how the decoherence suppression can be degraded with unavoidable delay times, by analyzing non-Markovian average dynamics. We present analytical expressions for the average dynamics and numerically analyze the effects of the delayed feedback for a coherently driven two-level system, coupled to a bosonic bath via a Hermitian coupling operator. We also find that, when the qubit's unitary dynamics does not commute with the measurement and feedback controls, the decoherence rate can be either suppressed or amplified, depending on the delay time. 
			\end{abstract}
	
	\maketitle
	
	
	\section{\label{sec:Introduction}Introduction}
	
	Decoherence for an open quantum system arises when the quantum system of interest interacts with its environment, resulting in the system losing its coherence properties and degrading its abilities to perform useful tasks for quantum technology~\cite{BookCarmichael,Div1998,zur2003}. To mitigate decoherence, there are techniques that have been proposed, such as 
	error correcting codes~\cite{BookQEC,Ter2015,Cam2017},
dynamical decoupling~\cite{Sup:DD1,Sup:DD2, Sup:DD3, Sup:DD4}, and measurement-feedback controls~\cite{Wiseman1993,BookWiseman,feedback1,Sup:FB1}. Most of the existing techniques consist of two main components; one is the part of collecting information (or knowledge) about the system via measurements, and another part is controlling the system to mitigate the decoherence effects, based on the collected information from the measurements. However, it was recently proposed that the decoherence cancellation was possible by \emph{only measuring the environmental noise affecting the system}, e.g., using  the so-called ``no-knowledge'' measurement~\cite{PhysRevLett.113.020407}. In the protocol, measurement results contain no information about the system and the decoherence effect \emph{could} be completely suppressed using feedback control based on the information of the environment noise.

	
	The no-knowledge quantum feedback~\cite{PhysRevLett.113.020407} was proposed as a decoherence cancellation protocol for a quantum system continuously coupled to a Markovian bosonic bath (environment). The protocol requires no need of any prior state filtering~\cite{Belavkin1992,BouVHan2007}, or knowledge about the measured system state. This is possible through a carefully chosen measurement setting, such that its measurement record is proportional to a Gaussian white noise $y(t) \propto \xi(t)$ (e.g., using a homodyne detection with a local oscillator phase $\pi/2$), and the feedback control is a simple function of the record. The no-knowledge quantum feedback was shown to completely cancel the backaction from the measurement. Once the backaction is perfectly reversed for all individual noise realizations, the decoherence on the system's average dynamics can be completely suppressed. The protocol can be applied in the case where the system-bath coupling operators are Hermitian, but can also be adapted for non-Hermitian coupling by adding extra conjugate channels (e.g., using gain and loss channels for bosonic baths) and mixing output records using beam splitters~\cite{PhysRevLett.113.020407,estimate, Carvalho_2011}.
	

	However, the analysis of the no-knowledge feedback so far has been shown for ideal situations, with a strong assumption that the measurement backaction on the system can be \emph{instantaneously} reversed by the feedback control. A more realistic model should take into account that, in any experiment, measurement processes take time and the feedback process occurs at a finite time after the measurement. From this more general model, one can investigate robustness of the decoherence suppression to a finite delay time. Moreover, the instantaneous process can still be properly defined by taking a limit when the delay time goes to zero. As long as the feedback process occurs after the measurement, the time-causal order is still preserved~\cite{Wiseman1993,BookWiseman}.

	
	In this work, we investigate the no-knowledge feedback with delay time via analytical calculations for open quantum systems, supplemented by numerical simulations for qubit examples. In addition to the stochastic master equations (SMEs) in Stratonovich interpretation used in the original work~\cite{PhysRevLett.113.020407}, we analyze the SMEs, describing individual quantum trajectories with the feedback control~\cite{Wiseman1993}, in both It\^o and Stratonovich interpretations~\cite{BookOksendal,Gardiner}.  The analysis shows that the decoherence suppression is degraded with the delay times and the complete suppression can only be achieved with the ideal zero delay time. With finite delayed time, the correct SMEs should be derived by treating the stochastic elements in the feedback and measurement processes as two independent noises. 
		
	In order to analyze the effect of delay time systematically, we implement the discrete-time operational approach for continuous quantum trajectory with feedback delay~\cite{Saiphet2019} and derive analytical solutions (whenever possible) for the system's average dynamics.  The discrete-time operational approach for quantum trajectories has been used in numerical simulations~\cite{rouchon2015efficient,jordan2015fluores,GueWis2020} and in processing measurement records from experiments~\cite{siddiqi2016,Kater2013,Shay2016noncom}, in order to reduce numerical errors from a finite-time resolution. The technique also allows us to investigate the ordering of operations applied to system's state with explicit delay times. We consider three different cases where the system's dynamics includes: (a) only the measurement and feedback, (b) with additional commuting unitary dynamics, and (c) with additional non-commuting unitary dynamics. We then compare the analytical results with numerically simulated dynamics for an example of a coherently-driven qubit coupled to a bosonic bath. Our results explicitly show how much the decoherence suppression can be degraded, as the delay time $\tau$ increases, i.e., from the ideal case of full cancellation ($\tau \rightarrow 0$) to the other limiting case of the pure decoherence (no feedback, or $\tau \rightarrow \infty$). We also show that, for the case (c), when the delay time is comparable to the time scale of the Hamiltonian evolution of the system, the decoherence rate can be either suppressed or amplified. By setting the delay time to be half integers of the qubit's Rabi period, the decoherence effect can even be worse than the case with no feedback at all. This similar effect has been recently found in a cavity-QED system with a time-delayed coherent feedback~\cite{Crow2020}.
	
	
	
	 This paper is organized as follows. In Section~\ref{sec:NK}, we briefly review the SMEs for homodyne measurement and the no-knowledge measurement with feedback control. We show that a contradiction in deriving the SMEs can arise from misinterpreting the stochastic feedback. In Section~\ref{sec:ImplementDelay}, we introduce the time-delayed feedback for the no-knowledge measurement and present our results for SMEs with finite delay time $\tau$. We then implement the time-discrete operation for quantum trajectories in Section~\ref{sec:Result}, derive the system's average dynamics, and show comparisons with numerical simulations for the qubit examples. The conclusion is in Section~\ref{sec:Conslusion} and a derivation for the time-delayed feedback SMEs is shown in Appendix~\ref{sec:App}.

	\section{\label{sec:NK}Perfect No-Knowledge Measurement and Feedback}

	We first briefly review the decoherence and stochastic master equations for homodyne detections, and then discuss no-knowledge quantum feedback with delay times. Let us first consider a quantum system coupled to $M$ bosonic baths via Lindblad operators $\hat c_j$ for $j = 1,2,..., M$ under the strong Markov assumption~\cite{LiLi2018}. For the case when there is no measurement on the bath, or measurement results are unknown, the system's dynamics is given by the Lindblad master equation (ME)~\cite{Lind1976},
	\begin{align}\label{decay}
	\partial_t \rho(t) &= {\cal L}\rho(t) \equiv  -i[\hat{H}, \rho(t)] + \sum_{j=1}^M  \mathcal{D}[\hat{c}_j]\rho(t),
	\end{align}
	where $\rho(t)$ is a quantum state matrix of the system. The superoperator $\mathcal{L}\bullet$ represents both the system's unitary dynamics (with the Hamiltonian $\hat{H}$) and the decoherence effect from the system-bath coupling via the Lindblad operators, where ${\cal D}[\hat{c}_j]\bullet \equiv \hat{c}_j\bullet \hat{c}_j^\dagger - \tfrac{1}{2}(\hat{c}_j^\dagger\hat{c}_j \bullet + \bullet \hat{c}_j^\dagger\hat{c}_j)$. The solution of the unconditional evolution Eq.~\eqref{decay} at any time $t$ is given by $\rho(t)= e^{\mathcal{L}t}\rho_0$ for an initial condition $\rho(t=0) = \rho_0$.

	However, when measurements are performed on the baths and their results are known to us (as an observer), the system's state dynamics can be conditioned on the measurement results, described by the quantum trajectory theory~\cite{BookCarmichael,BookWiseman,BookJacobs}. For simplicity, let us assume $M=1$ ($\hat c_1 = \hat L e^{i \theta}$) and the bath's detection is performed via the homodyne measurement with a phase $\theta$, where a measurement record $y_{\theta}(t)$ is acquired for time $t \in [0, T)$. The quantum system's dynamics conditioned on the homodyne record is described by the diffusive-type SMEs, which can be written in the It\^o and Stratonovich formulations. We first write the Stratonovich SME for the quantum state under the homodyne detection with a measurement efficiency $\eta$,
	\begin{equation}
	\partial_{t}{\rho}(t) = \mathcal{L}{\rho}(t) + \sqrt{\eta}\,y_{\theta}(t)\mathcal{H}[\hat{L}e^{i\theta}]{\rho}(t)  - \frac{\eta}{2}\mathcal{A}^{2} [\hat{L}e^{i\theta}]{\rho}(t), \label{Stratonovich}
	\end{equation}
	where we have defined the superoperators $\mathcal{H}[\hat{c}]\bullet = \hat{c}\bullet + \bullet\hat{c}^{\dagger} - {\rm Tr}(\hat{c}\bullet + \bullet\hat{c}^{\dagger})\bullet$ and $\mathcal{A}^{2}[\hat{c}]\bullet= \bar{\mathcal{A}}^2[\hat c]\bullet- {\rm Tr}(\bar{\mathcal{A}}^2[\hat c]\bullet)\bullet$ using $\bar{\mathcal{A}}[\hat c]  = \hat c \bullet + \bullet \hat c^\dagger$. The last two terms in Eq.~\eqref{Stratonovich} represent the stochastic term due to the measurement backaction and the Stratonovich correction term, respectively. The measurement signal can also be written as $y_{\theta}(t) = \sqrt{\eta}{\rm Tr}[( \hat{L}e^{i\theta} + \hat{L}^{\dagger}e^{-i\theta}) \rho(t) ] + \xi(t)$, which contains the information about the system and a stochastic Gaussian white noise $\xi(t)$. For completeness, we also note that the Stratonovich SME in Eq.~\eqref{Stratonovich} is equivalent to the following It\^o SME
	\begin{align}\label{Ito}
	\dd \rho(t) = {\cal L}\rho(t) \dt + \sqrt{\eta}\, \xi(t) {\cal H}[\hat{L} e^{i \theta}]\rho(t) \dt,
	\end{align}
	where we have used ``$\dd \rho$'' on the left-hand side to indicate the explicit differential in the It\^o formalism~\cite{BookWiseman}. In Eq.~\eqref{Ito}, we wrote the stochastic term using the Gaussian white noise as defined before, but note that $\xi(t)\dt$ has the same statistical properties as the Wiener increment $\dd W(t)$~\cite{BookOksendal,Gardiner}. The above SMEs, after averaging over all possible measurement record realizations, should both agree with the Lindblad decay evolution in Eq.~\eqref{decay}. This averaging over the records is equivalent to tracing over the bath's degree of freedom. 
		
	The no-knowledge measurement in Ref.~\cite{PhysRevLett.113.020407} was introduced with a set of conditions, such that the measurement device only measures the noise of the environment. The conditions are that $\hat{L}$ must be Hermitian ($\hat L = \hat L^\dagger$) and that the homodyne phase must be $\theta = \pi/2$. These lead to
	\begin{align}
	\hat{L}e^{i\pi/2} + \hat{L}^{\dagger}e^{-i\pi/2}  = 0,
	\end{align}
	and the measurement signal is of the form, 
	\begin{align}\label{NKrecord}
	y_{\pi /2}(t) = \xi(t),
	\end{align}
	which contains no information about the measured state, only about the stochastic noise at time $t$. 

	Following the original proposal that used Stratonovich SMEs and assuming a perfect homodyne measurement, $\eta = 1$, we can obtain the SMEs by replacing the measurement signal with the white noise, i.e., substituting Eq.~\eqref{NKrecord} in Eq.~\eqref{Stratonovich}. Given that $\hat{L} = \hat{L}^\dagger$, one obtains $\tfrac{1}{2}{\cal A}^2[\hat{L}e^{i\pi/2}]\rho = {\cal D}[\hat L e^{i\pi/2}]\rho$ that cancels the decoherence term in ${\cal L}\rho$ in Eq.~$\eqref{Stratonovich}$. The Stratonovich SME for the no-knowledge measurement becomes~\cite{PhysRevLett.113.020407},
	\begin{equation}
	\partial_{t} {\rho}(t) = -i [\hat{H} - y_{\pi /2}(t)\hat{L}, {\rho}(t)],  \label{NKeq}
	\end{equation}
 which clearly shows that the measurement backaction on the system is unitary and therefore \emph{could} be invertible. This was the essence of the decoherence cancellation in the proposal, where the system's Hamiltonian was engineered so that the measurement backaction could be completely cancelled, i.e., by replacing $\hat{H}$ with $\hat{H}_{\rm eff}(t) = \hat{H} + y_{\pi/2}(t) \hat{L}$. With this modification, the dynamics in Eq.~\eqref{NKeq} is reduced to the system's bare unitary dynamics,
	\begin{align}
	\partial_{t} {\rho}(t) & =-i [\hat{H}_{\rm eff}(t) - y_{\pi /2}(t)\hat{L}, {\rho}(t)],\nonumber \\
	& =  -i [\hat{H}, {\rho}(t)], \label{PerfNK}
	\end{align}
	which is the desired dynamics unaffected by the decoherence. The engineered Hamiltonian can be considered as a result of adding a \emph{feedback} Hamiltonian, $\hat{H}_{\rm F}(t)= y_{\pi/2}(t) \hat{L}$, to the system at time $t$, using the measured signal $y_{\pi/2}(t)$. This is the perfect measurement-feedback protocol with no delay time or latency. 

	
 To motivate the use of finite delay times, we will show that the simple replacement of $\hat{H}$ with $\hat{H}_{\rm eff}(t) = \hat{H}+\hat{H}_F(t)$ in Eq.~\eqref{PerfNK} (also in Ref.~\cite{PhysRevLett.113.020407}) leads to the correct instantaneous limit for the Stratonovich SME, but it fails to reproduce a correct SME when using the It{\^o} formulation. Indeed, applying the no-knowledge conditions and replacing $\hat{H}$ with $\hat{H} + \hat{H}_{\rm F}(t)$ to the It\^o SME in Eq.~\eqref{Ito}, we obtain
	\begin{equation}
	\dd{\rho}(t) = -i [\hat{H}, {\rho}(t)]\dt + \mathcal{D}[\hat{L}]{\rho}(t)\dt = {\cal L}\rho(t) \dt. \label{ItoNK}
	\end{equation}
	That is, the resulting dynamics is not the simple bare unitary dynamics as in Eq.~\eqref{PerfNK}, but there is a decoherence effect on the system. This contradiction emerges because, by simply adding the feedback Hamiltonian $\hat{H}_{\rm F}(t)$ in the It\^o SME in Eq.~\eqref{ItoNK}, one had mistakenly assumed that the feedback was not stochastic. This is not true as the feedback Hamiltonian is a function of the record $y_{\pi/2}(t)$, and the stochastic differential equations should be treated carefully when it involves noise terms that are not differentiable~\cite{BookOksendal,Gardiner}. By including finite delay times into the problem, we can explicitly determine whether the noise from measurements should be correlated or independent from the noise in the feedback term. In the next Section, we will show how to reconcile these two formulations by appropriately treating the effect of delay times in the SMEs for the no-knowledge feedback.

	\section{\label{sec:ImplementDelay}No-knowledge Measurement and Feedback with delay time}

\begin{figure}[t]
    \centering
    \includegraphics[width=0.6\linewidth]{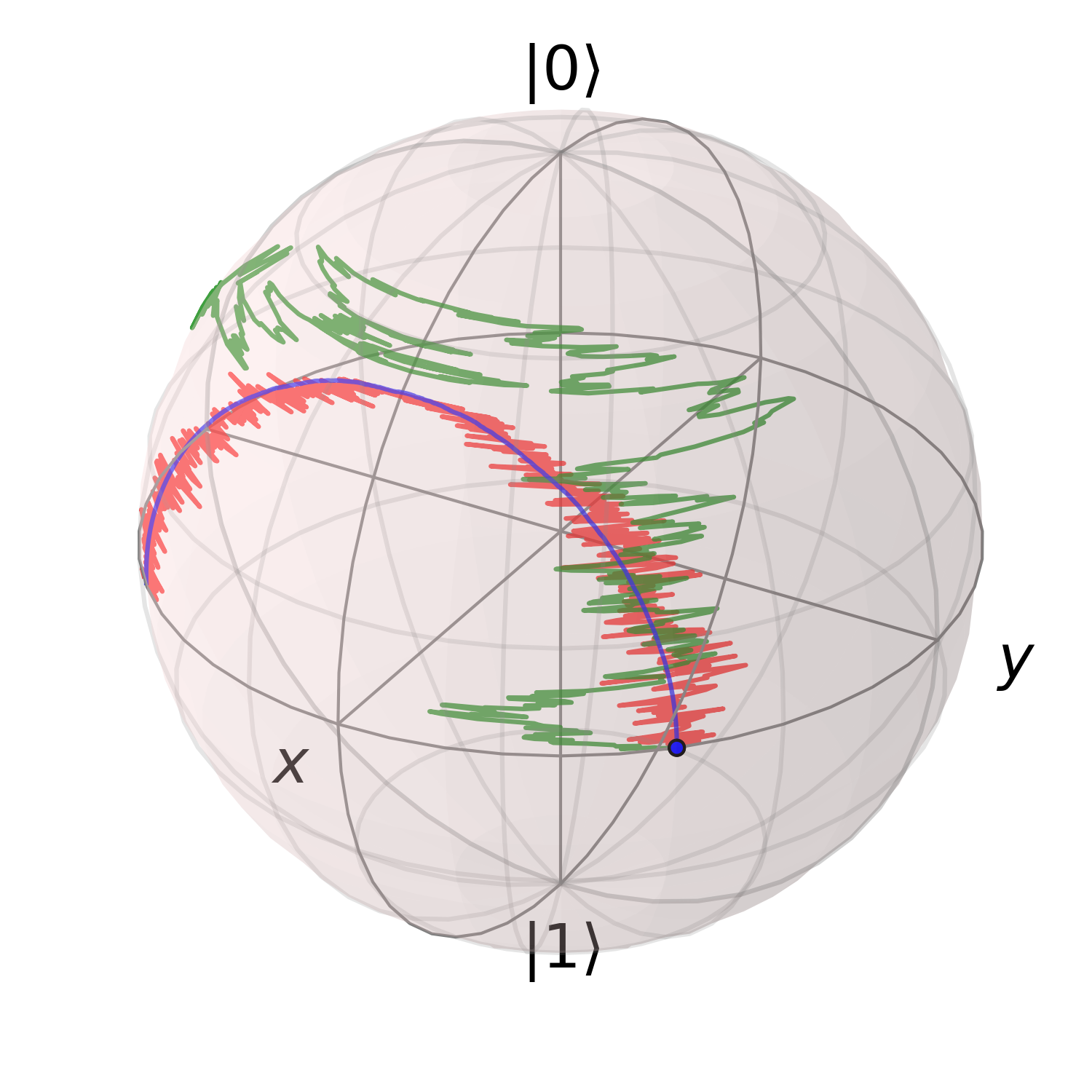}
    \caption{Individual realizations, shown on the Bloch sphere, for the qubit example under the no-knowledge measurement and feedback protocols. For the perfect case with no delay time, $\tau = 0$, the qubit evolves with the bare unitary evolution, described by the smooth blue curve. The red and green curves represent qubit's trajectories for the delay times $\tau = 10^{-3} T_\Omega$ and $\tau = 0.1 T_\Omega$, respectively. The initial state $\rho_{0} = \frac{1}{2}(\mathbbm{1} + \frac{1}{\sqrt{2}} (\hat{\sigma}_{x} + \hat{\sigma}_{y}))$ is indicated by the blue dot. The parameters are $\Omega = 2\pi T_\Omega^{-1}$ and $\ddt = 10^{-3} T_\Omega$.}
    \label{fig:Bloch_Intro}
\end{figure}	 
	
		
	As preluded in the previous section, we then need to consider the measurement process as a separate process from the feedback control. We use a discrete-time operational method to compute the system state dynamics. The dynamics is decomposed into discrete-time operations with the time resolution determined by a small, but finite, timestep $\ddt$. This technique allows us to investigate the effect of ordering of operations applied to the system's state and investigate the delay time by explicitly specifying when feedback operations occur.  The operational technique also guarantees that the quantum state is normalized and positive semi-definite at every time step in the numerical simulation. Moreover, by expanding such operations to first order in $\ddt$ and taking the continuum limit $\ddt \rightarrow \dt$, where $\dt$ is an infinitesimal time, one can get back the usual SMEs.
	
	For the system with no-knowledge feedback, there are three types of dynamics that describe the system's state during a timestep $\ddt$: (a) the system's unitary evolution, (b) the measurement backaction, and (c) the dynamics from the feedback control. These can be written mathematically as three separate operations. The system's unitary dynamics is described by $\mathcal{U}_{\ddt}[\bullet] = \hat{U}\bullet\hat{U}^{\dagger}$, where $\hat{U} = \exp(-i\hat{H}\ddt)$ is a unitary operator. The measurement backaction, from the no-knowledge measurement acquiring the noise $y_{\pi/2}(t) = \xi(t)$ between time $t$ and $t+\ddt$, is described by a measurement operation $\mathcal{M}_{t}[\bullet] = \hat{M}_{t}\bullet\hat{M}^{\dagger}_{t}$, where
	\begin{align}
	\hat M_t = & \, \,  \, \,  \exp[ i\xi(t)\hat{L}\ddt ],
	\end{align}
	is the measurement operator. We note that this operator can also be obtained from the conventional form of the Kraus operator for diffusive continuous measurement~\cite{BookCarmichael,BookWiseman}, $\hat M_t = \hat{1}-\frac{1}{2}\hat{c}^\dagger\hat{c}\ddt+\hat{c}\, y_{\pi/2}(t) \ddt + {\cal O}(\ddt^2)$, for the Lindblad operator $\hat c = \hat L e^{i \pi/2} = i \hat L$ and the record $y_{\pi/2}(t) = \xi(t)$, applying the It\^o rule $\xi(t)^2 \ddt^2 \sim \ddt$~\cite{BookOksendal,Gardiner}.

	For the no-feedback case, we can express the system's state dynamics from time $t$ to $t+\ddt$ as
	\begin{align}\label{eq-MU}
	\brho(t+\ddt) = \hat{U}\hat{M}_{t}{\brho}(t) \hat{M}^{\dagger}_{t}\hat{U}^{\dagger},
	\end{align}
	where $\brho(t)$ is an unnormalized state of the system at time $t$. One can show that Eq.~\eqref{eq-MU} is equivalent to the Stratonovich SME in Eq.~\eqref{NKeq} and the It\^o SME in Eq.~\eqref{ItoNK}. For the former, the proof can be done by expanding terms in Eq.~\eqref{eq-MU} to the first order in $\ddt$ and then taking the time-continuum limit $\ddt \rightarrow \dt$. For the latter, the equivalence can be shown by expanding Eq.~\eqref{eq-MU} to the order containing $\xi(t)^2\ddt^2$ and applying the It\^o rule. The equivalence, however, is valid only when $\ddt$ is small enough in order to ignore the contribution from higher-order terms. Moreover, the ordering of $\hat{M}_{t}$ and $\hat{U}$ here is irrelevant, because the error from two different orderings, i.e., $\hat{M}_t\hat{U}\bullet \hat{U}^{\dagger}\hat{M}^{\dagger}_{t} = \hat{U}\hat{M}_t\bullet \hat{M}^{\dagger}_t\hat{U}^{\dagger} + {\cal O}(\ddt^{2})$, from the non-commuting $\hat{U}$ and $\hat{M}_{t}$, is of the order $\ddt^{2}$, which is not included~\cite{heinosaari_ziman_2013, Varga}.
	
	Let us now consider the feedback process applied to the system with a delay time $\tau$. Following the intuition in Eq.~\eqref{NKeq}, with the modified Hamiltonian $\hat{H}_{\rm eff}(t)$, the feedback operation at time $t$ can be defined as ${\cal F}_t[\bullet] = \hat{F}_t \bullet \hat{F}_t^\dagger$ where $\hat{F}_{t} = \hat{M}_{t-\tau}^{\dagger} = \exp[-i \xi(t-\tau)\hat{L}\ddt]$, using the measurement signal acquired at time $t-\tau$. We can therefore express the system's dynamics, when the feedback is turned on,
	\begin{align}\label{eq-MUF}
	\brho(t+\ddt) = \hat{F}_t\hat{U}\hat{M}_t{\brho}(t)\hat{M}_t^{\dagger}\hat{U}^{\dagger}\hat{F}_t^{\dagger},
	\end{align}
	where we explicitly put the feedback operator \emph{after} the measurement operator~\cite{Wiseman1993}. This is so that the feedback process still occurs after the measurement, even in the limit of $\tau \rightarrow 0$. 
	
	We can then derive (see Appendix~\ref{sec:App} for the derivation) the corresponding SMEs for the no-knowledge measurement with time-delayed feedback Eq.~\eqref{eq-MUF}. For the Stratonovich interpretation, we expand Eq.~\eqref{eq-MUF} to first order in $\ddt$~\cite{wongzakai1965} and take $\ddt \rightarrow \dt$ to get
	\begin{align}
	\partial_t {\rho}(t) = -i[\hat{H}, {\rho}(t)] - i[\hat{L}, {\rho}(t)]\{\xi(t- \tau) - \xi(t)\}, \label{E1}   
	\end{align}
	as the Stratonovich SME. For the It\^o interpretation, one has to be careful at treating the two noises, $\xi(t)$ and $\xi(t-\tau)$, as independent, when $\tau \ne 0$. That is, the It\^o rule has be to applied separately to both noises. By expanding Eq.~\eqref{eq-MUF} to first order in $\ddt$, $\xi(t)^2\ddt^2 = \ddt$, and $\xi(t-\tau)^2\ddt^2 = \ddt$ and taking the continuum limit, we obtain
	\begin{align}
	\dd{\rho}(t) =& -i[\hat{H}, {\rho}(t)]\dt + 2\mathcal{D}[\hat{L}]\rho(t)\dt \nonumber \\
	&  - i[\hat{L}, {\rho(t)}]\{\xi(t- \tau) - \xi(t)\}\dt,  \label{E2}
	\end{align}
as the It\^o SME for the delayed feedback. Eqs.~\eqref{E1} and \eqref{E2} are non-Markovian SMEs because of the delay time.

Now, a question arises as whether we can obtain no-delay SMEs by simply taking $\tau =0$ in the above equations, even though it is not practical in real experiments? The answer is still no. This is because, with  zero delay time, the two noises in the feedback operator and the measurement operator are exactly the same and strongly correlated. That is, the cross term, $\lim_{\tau \rightarrow 0} \xi(t) \xi(t-\tau) \ddt^2 = \ddt$, should now be taken into account. Therefore, one needs to use Eq.~\eqref{eq-MUF}, taking $\tau =0$ and expanding terms to first order in $\ddt$ and $\xi(t)^2\ddt^2 = \ddt$. We then get
		\begin{align}
	\dd{\rho}(t) &= -i[\hat{H}, {\rho}(t)]\dt ,
	\end{align}
as the correct It\^o SME for the no-delay feedback. This is in agreement with the bare unitary dynamics in the Stratonovich interpretation in Eq.~\eqref{PerfNK}.

	\section{\label{sec:Result}Average dynamics and example of qubit measurement}
	
	For the ideal, but unphysical, case of the instantaneous feedback, the quantum state dynamics is simply described by the bare unitary dynamics as in Eq.~\eqref{PerfNK}. However, for the feedback with delay time, the system's evolution can become fluctuating from the effect of noise that is not perfectly cancelled. Examples are presented in Fig.~\ref{fig:Bloch_Intro} for a single two-level system coupled to a bosonic bath, where the perfect cancellation results in the smooth rotation (blue curve) and the delayed feedback results in the fluctuating curves (red and green curves are for small and large delay times, respectively). Because of the fluctuation, it is difficult to analyze the effect of delay times for individual trajectories. Therefore, we instead investigate how the delay times affect the system's dynamics on average. 
	
	In this section, we separate the investigation into three cases, showing the analytical solutions whenever possible and numerical results for a single qubit example. The first case is when the unitary dynamics is neglected and the system is evolved with only the measurement and the feedback processes. The second case is when there is the unitary dynamics, but its Hamiltonian commutes with the system-environment coupling operator. Finally, the third case is when the Hamiltonian does not commute with the coupling operator. 
	
	For the numerical results throughout this paper, we use a single qubit coupled to a Markovian bosonic environment through a Hermitian coupling (Lindblad) operator $\hat{L} = \sqrt{\gamma} \hat \sigma_z$. The qubit's unitary dynamics is the Rabi oscillation described by $\hat{H} = \tfrac{\Omega}{2} \hat{\sigma}_{j}$ where $\hat{\sigma}_{j}$ is a Pauli matrix which determines the axis of the rotation. For the numerical simulation, we generate qubit's trajectories using the operational forms in Eqs.~\eqref{eq-MU} and \eqref{eq-MUF} with random white noises (Gaussian distribution with zero mean and a standard deviation $1/\sqrt{\ddt}$) using Python programming. The qubit's initial state is set to $\rho_{0} = \frac{1}{2}(\mathbbm{1} + \frac{1}{\sqrt{2}} (\hat{\sigma}_{x} + \hat{\sigma}_{y}))$ for simplicity. The time step $\ddt$ is chosen to be small enough to satisfy $\ddt \ll T_\Omega$, where $T_\Omega = 2\pi/\Omega$ is the Rabi period.
	

	\begin{figure}
	\centering
	\includegraphics[width=1.0\linewidth]{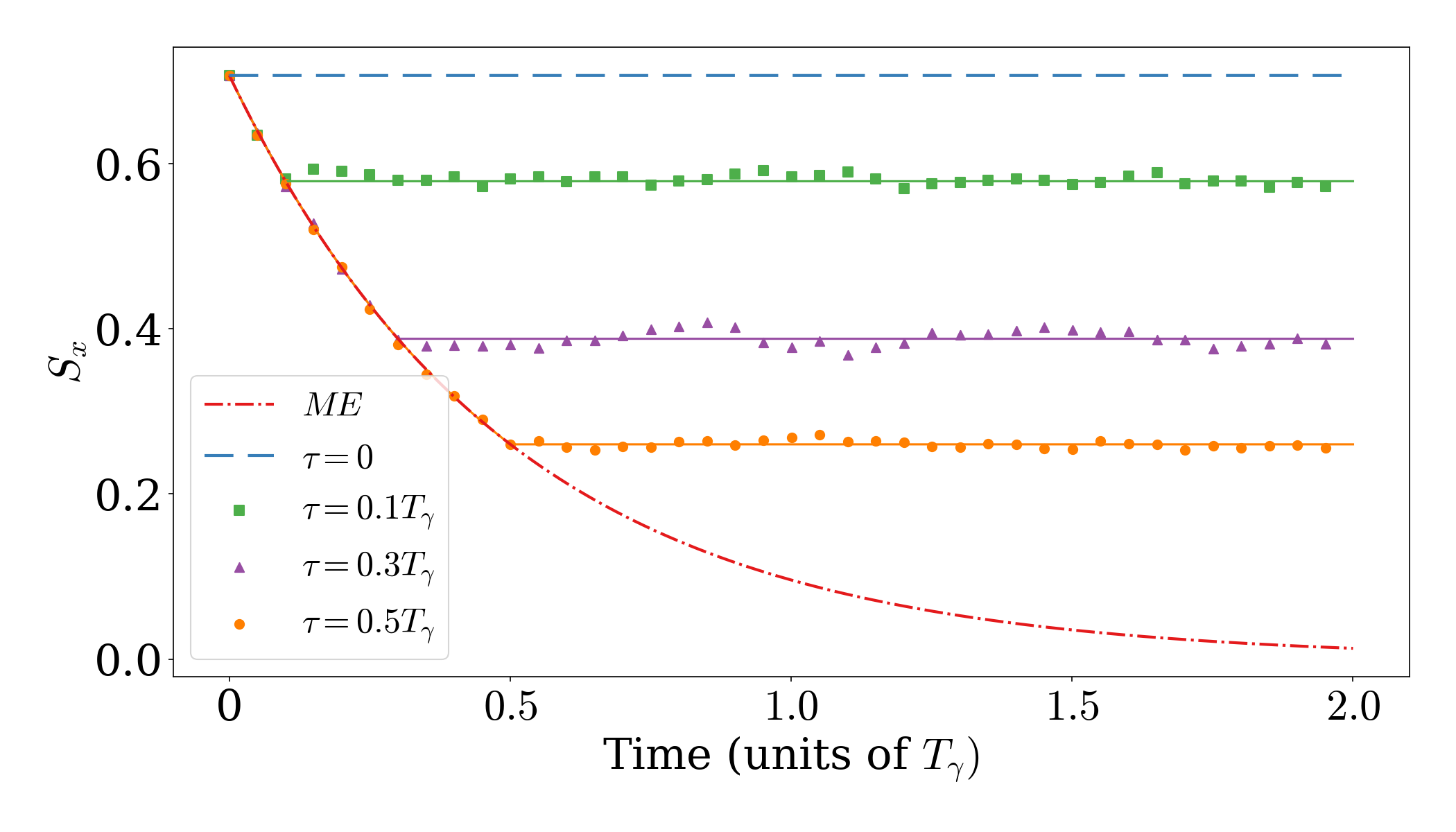}
	\caption[Delayed feedback with no Rabi oscillation.]{Average dynamics for the feedback process without the unitary evolution. The plot shows average qubit's dynamics for the $x$-coordinate, $S_x(t) = {\rm Tr}(\rho_{\rm av}(t) \hat\sigma_x)$, for the Lindblad master equation (ME) and $\tau = \alpha T_\gamma$, where $T_\gamma = 1/\gamma$ is the dephasing time and $\alpha = 0.1$, $0.3$, and $0.5$. The decoherence cancellation in this case results in the stabilization of the system's state, which occurs after the feedback process is present. The analytical solutions Eq.~\eqref{case11} are shown in solid lines and the numerically simulated results are shown as colored dots. Since the measurement backaction is to simply rotate the qubit around the $z$-axis, the averaging effect is similar for both in the $x$ and $y$ coordinates and the dynamics in $z$-coordinate is constant in time. }
	\label{Fig1:case1}
\end{figure}
	
	\subsection{\label{subsec:1st}Feedback process without unitary dynamics}
	
	For the first case, we assume no qubit's unitary evolution, so that we can solely investigate effects of the measurement and feedback cancellation with delay time. During the time before the feedback occurs, i.e., $t < \tau$, the system's dynamics is a result of only the measurement backaction, which leads to an average dynamics described by the complete decoherence in Eq.~\eqref{decay}.
	Once the feedback kicks in, when $t \ge \tau$, the system's dynamics includes the evolution from the feedback operator $\hat{F}_{t} = \hat{M}_{t-\tau}^\dagger =  \exp\{-i \xi(t-\tau)\hat{L}\ddt\}$. Since the feedback operator always commutes with the measurement operator, $[\hat{M}_t, \hat{F}_{t'} ] = 0$, for any times $t,t'$, as they are inverses of each other, any added feedback operators can subsequently cancel the measurement operators with the same measurement results at $\tau$-time earlier, i.e., $\hat{F}_{t} \hat{M}_{t-\tau} = \hat 1$.

	Therefore, using the update equation in Eq.~\eqref{eq-MUF}, one can find that there is always effectively a fixed number ($m=\tau/\ddt$) of the measurement operators that cannot be cancelled by the feedback operators, for any time $t \ge \tau$. We can obtain an analytic solution for the average evolution, 
	\begin{equation}
	\rho_{\rm av}(t) = 
	\begin{cases}
	e^{\mathcal{D}[\hat{L}]t} \rho_{0}, &      \text{      t $ < \tau $ }\\
	e^{\mathcal{D}[\hat{L}]\tau} \rho_{0}, &	     \text{    t $ \geq \tau $ }
	\end{cases}. \label{case11}
	\end{equation}
	This describes the usual Lindblad decay (without the unitary dynamics) for $t < \tau$, and the dynamics after $t = \tau$ is frozen at the state $\rho_{\rm av}(\tau) =e^{\mathcal{D}[\hat{L}]\tau} \rho_{0}$. For the instantaneous feedback, the state should be stabilized at the initial state $\rho_0$. As a result, the decoherence cancellation is not perfect for a finite delay time, where the stabilized state is degraded from the initial state. 
	
We apply the above result to the qubit example where $\hat L =\sqrt{\gamma} \hat\sigma_z$. The steady-state fidelity between the initial state and the degraded stabilized state can also be found analytically for the pure initial state $\rho_0$,
	\begin{align}
	F & \equiv {\rm Tr}[\rho_0\, \rho_{\rm av}(\infty)] \nonumber \\
	& = \frac{1}{2}\left[ (1+S_z(0)^2) + (1-S_z(0)^2)e^{-2\gamma\tau} \right],
	\end{align}
where $S_z(0) = {\rm Tr}(\rho_0 \hat\sigma_z)$.	We show, in Fig.~\ref{Fig1:case1}, the dynamics of the qubit example from numerically simulating $5\times 10^3$ qubit trajectories, for $\tau = \alpha T_\gamma$, where $T_\gamma = 1/\gamma$ is the dephasing time and $\alpha = 0.1$, $0.3$, and $0.5$.

	\begin{figure}[t]
	\centering
	\includegraphics[width=1.0\linewidth]{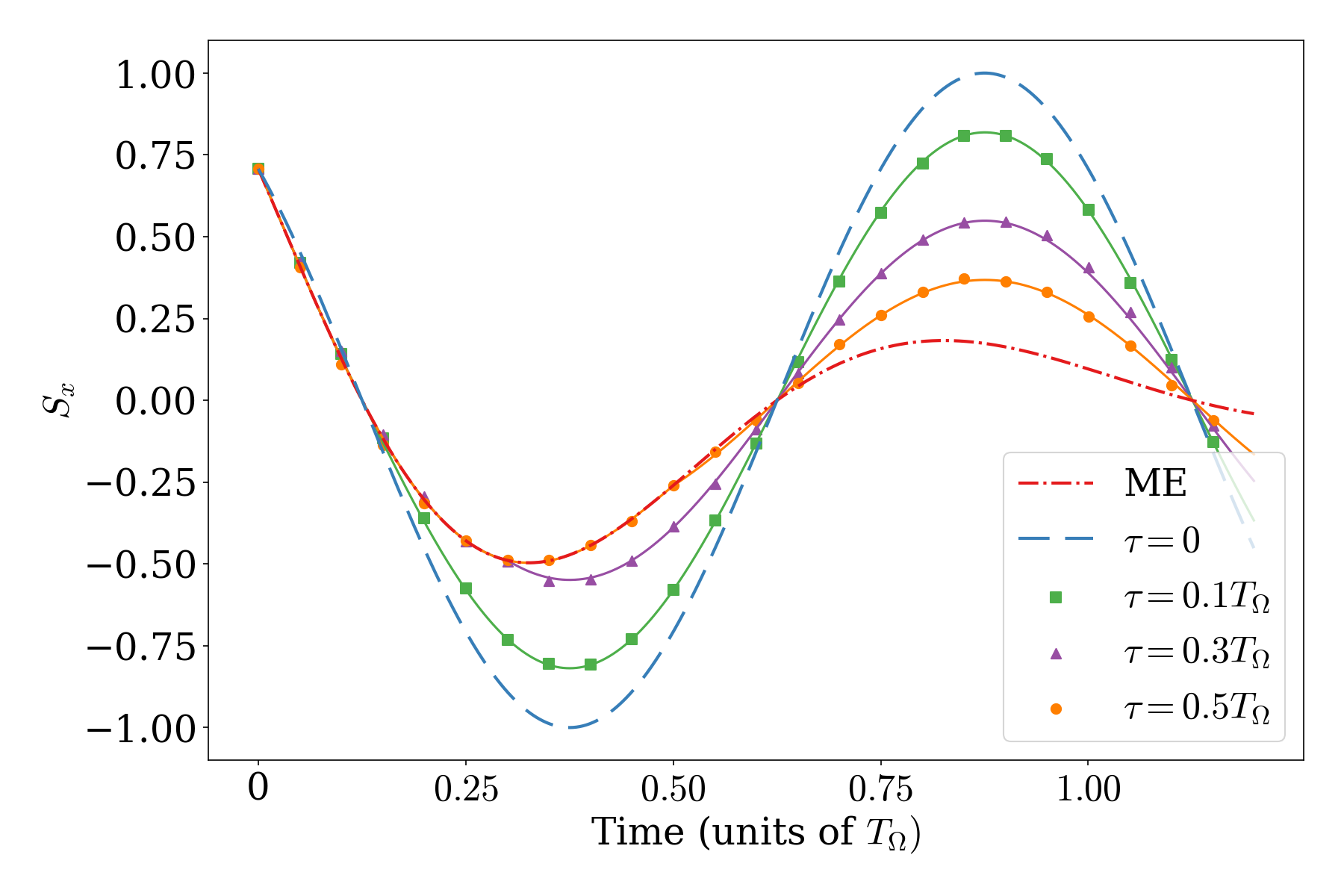}
	\caption[Delayed feedback with Rabi oscillation.]{Average dynamics for the feedback process with commuting unitary evolution. The plot shows average qubit's dynamics for the $x$ and $z$ coordinates, $S_x = {\rm Tr}(\rho_{\rm av}(t) \hat\sigma_x)$ and $S_z = {\rm Tr}(\rho_{\rm av}(t) \hat\sigma_z)$, for the master equation (ME) and $\tau = \alpha T_{\Omega}/2$ where $\alpha = 0.2, 0.6, 1.0$. The qubit's state average dynamics follows the Lindblad ME decay for $t < \tau$ and deviates from it after that time, showing that the feedback can restore the amplitude of Rabi oscillation inverse proportional to the delay. Analytical results from Eq.$\eqref{case22}$ are plotted in solid curves with numerical results in dots. The plot of $z$-coordinate will be used to compare with results from the non-commuting case.}
	\label{Fig1:case2}
\end{figure}
	
	\subsection{\label{subsec:2nd}Feedback process with commuting unitary dynamics}
	For the second case, we include the commuting unitary dynamics $\hat{U} = \exp(-i\hat{H}\ddt)$ to the system, given that $[\hat{H}, \hat{L}] = 0$. We show that the feedback operation can still cancel the measurement backaction in a similar manner as in the previous case. Let us introduce the time indices $j \in \{ 0,1, 2, ..., n\}$, where the initial time is $t_0 = 0$, an arbitrary time is $t_j = j\ddt$, a time of interest is $t = n \ddt$, and the delay time is $\tau =  \kappa \ddt$. We denote the measurement noise $\xi_j = \xi(t_j)$ as the noise acquired between $t_j$ and $t_{j+1}$. Therefore, the feedback operator applied to the system at time $t_j$ is a function of the measurement noise at time $t_j-\tau$, which is given by $\hat{F}_{t_j} = \hat{M}^\dagger_{t_j-\tau} = \exp(- i \xi_{j-\kappa}\hat{L}  \ddt$).

	During the first $\kappa$ time steps (or when $t < \tau$), there is no feedback on the system and the system evolves only via the unitary dynamics and the measurement backaction, i.e., ${\brho}(t+\ddt) = \hat{U}\hat{M}_t{\brho}(t) \hat{M}_t^\dagger \hat{U}^\dagger$. The average dynamics for this part is therefore the Lindblad decay $\rho_{\rm av}(t) = e^{{\cal L}t}\rho_0$ for $t< \tau$. At the delay time $\tau = \kappa \ddt$, the feedback process starts by feeding in the dynamics that is a function of the noise at $\tau$-time earlier, leading to a different dynamics described by ${\brho}(t+\ddt) = \hat{M}^\dagger_{t-\tau}\hat{U}\hat{M}_t{\brho}(t) \hat{M}_t^\dagger \hat{U}^\dagger \hat{M}_{t-\tau}$ for $t \ge \tau$.

\begin{figure*}[t]
    \begin{center}
    \includegraphics[width=0.9\linewidth]{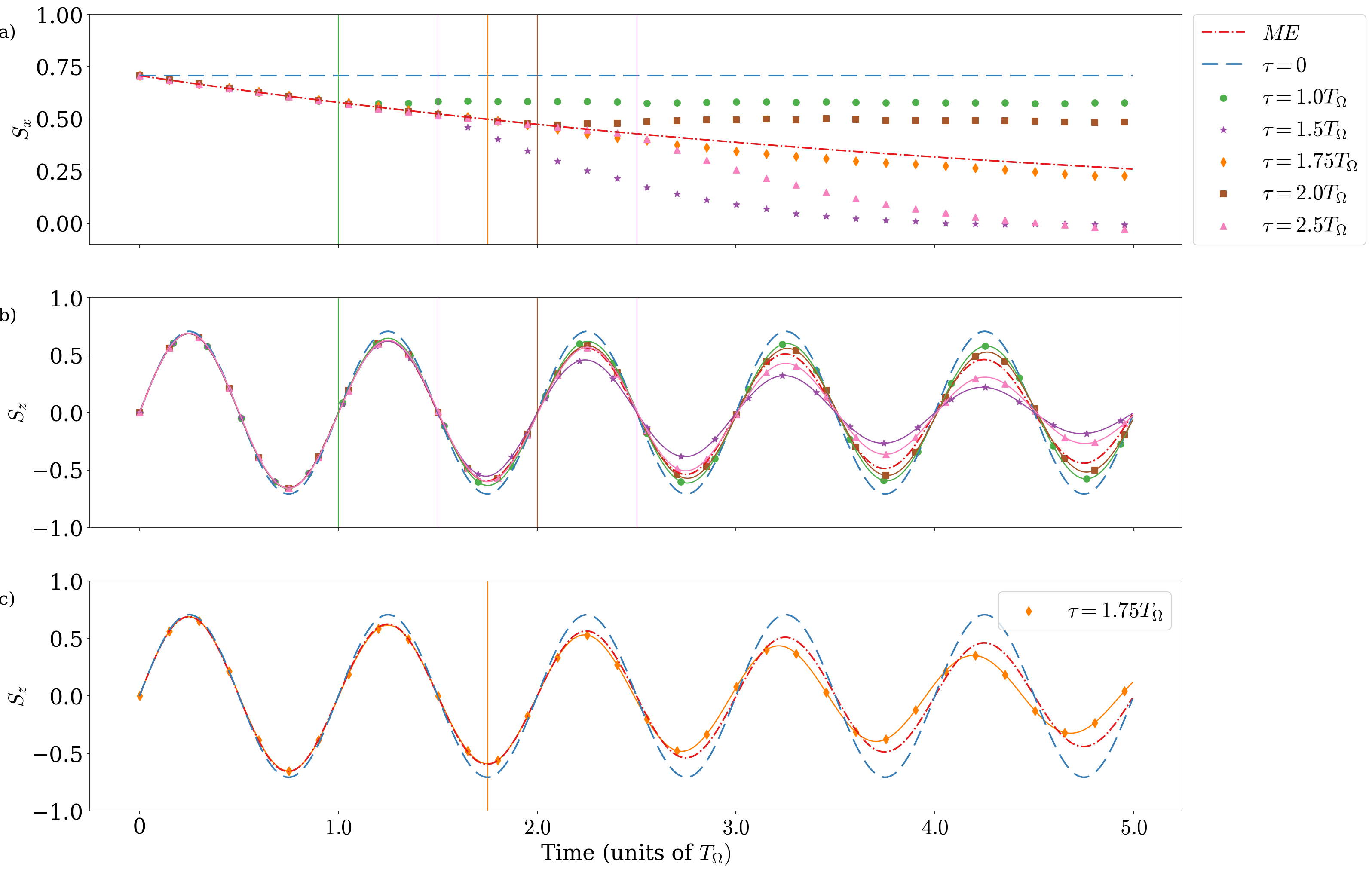}
    \end{center}
    \caption[Delayed feedback with non-commuting $\hat{H}$]{Average dynamics of the qubit for the no-knowledge feedback process with non-commuting unitary evolution, for the master equation (ME) and $\tau = \alpha T_{\Omega}/2$ where $\alpha = 2.0$, $3.0$, $3.5$, $4.0$, and $5.0$. The Rabi oscillation is chosen to be around the $x$-axis with the frequency $\Omega$, where the measurement backaction and feedback kick the qubit in the $x$-$y$ plane. Panel (a) shows the averaged $x$-coordinate of the qubit $S_x(t) = {\rm Tr}[\rho_{\rm av}(t)\hat\sigma_x]$ for all chosen delay times, while panels (b) and (c) show the averaged $z$-coordinate $S_z(t) = {\rm Tr}[\rho_{\rm av}(t)\hat\sigma_z]$. The pure unitary evolution is shown as dashed blue curves and the Lindblad decay is shown as solid blue curves in all panels. The vertical lines represent the times at which the delayed feedback starts. Numerically simulated data are presented as colored markers for the in-phase ($\tau = k T_\Omega$) and the out-of-phase ($\tau = \tfrac{k}{2} T_\Omega$) conditions. We also show an intermediate case (the orange $\blacklozenge$-shape marker), where $\tau = 1.75T_\Omega$, separately in panel (c) for clarity. In this simulation, we use $\ddt = 10^{-4} T_\Omega$.}
	\label{Fig1:case3}
\end{figure*}
	
	Given that all operators commute, we can write the quantum state at any time $t = n\ddt \ge \tau$ conditioned on the measurement results and the feedback as
	\begin{eqnarray}
	\rho_{\rm c}(t) &=& (\hat{F}_{t-\ddt} \cdots \hat{F}_{\tau})(\hat{M}_{t-\ddt}\cdots \hat{M}_{0})\hat{U}^n \rho_{0} (\cdots)^\dagger  \nonumber \\
	&=& \mathcal{U}_\kappa^n \rho_{0}\, \mathcal{U}_\kappa^{n\dagger}, \label{UrhoU} 
	\end{eqnarray}
	where we have rearranged the operators so that the same types are grouped together. In the first line, we used $(\cdots )^\dagger$ to represent the adjoint of all operators on the left side of $\rho_0$. In the second line, we denoted a single operator $\mathcal{U}_\kappa^n$ for those operators, where the superscript and the subscript denote the time of interest $t = n\ddt$ and the delay time $\kappa\ddt$, respectively. The operator can be written explicitly as
	\begin{eqnarray}\label{eq-calU}
	\mathcal{U}_\kappa^n =&  \exp\left(-i\hat{L}\ddt \sum_{k=0}^{n-\kappa-1} \xi_{k}\right) \exp\left(i\hat{L}\ddt \sum_{k=0}^{n-1} \xi_{k}\right) \nonumber \\
	& \times \exp\left(-i\hat{H}n \ddt\right),
	\end{eqnarray}
	using the definition of noises as defined earlier. The first and the second exponential terms come from the feedback operators and the measurement operators, respectively. We can see that the feedback operations cancel some of the measurement backaction, leaving terms with noises at $\kappa$ times, which are $\xi_{n-\kappa}, \cdots, \xi_{n-1}$.

	Knowing that the noises all have Gaussian distribution, we can therefore solve for the ensemble average of the dynamics, for $t \ge \tau$, by averaging over all realizations of the uncanceled noises,
	\begin{eqnarray}\label{avedyn}
	\rho_{\rm av}(t)&=&  \int \!\! {\cal D}\xi \, \, {\cal P}(\xi_{n-\kappa}, \cdots, \xi_{n-1}) \, \mathcal{U}^{n}_{\kappa} \rho_{0} \mathcal{U}^{n\dagger}_{\kappa}.
	\end{eqnarray}
	where we have defined the measure for the integration as $\int{\cal D}\xi$ $=$ $\int \prod_{k=n-\kappa}^{n-1} \dd\xi_{k}$. The joint probability distribution of white noises is defined as 
	\begin{align}
	{\cal P}(\xi_a \cdots \xi_b)=\prod_{k=a}^b G(\xi_k; 0, 1/\sqrt{\ddt}),
	\end{align}
	that is, a product of Gaussian distributions with zero means and variances $1/\sqrt{\ddt}$. Since the noises $\xi_j$ are independent from each other, we can simplify the average in Eq.~\eqref{avedyn} by separating out $n-\kappa$ unitary operators and doing the integration over the $\kappa$ white noises, defined for simplicity as $\eta_1 \cdots \eta_\kappa$, and obtain
	\begin{eqnarray}
	\rho_{\rm av}(t) &=& \hat{U}^{n-\kappa}  \varrho(\tau)( \hat{U}^\dagger)^{n-\kappa} ,
	\end{eqnarray}
	where 
	\begin{align}\label{avedyn2}
	\varrho(\tau) = \!\! \int \!\! {\cal D}\eta \, \, {\cal P}(\eta_1, ..., \eta_\kappa) \hat{U} e^{i \hat{L} \ddt \eta_\kappa} \cdots \hat{U} e^{i \hat{L}\ddt \eta_1} \rho_0 (\cdots)^\dagger,
	\end{align}
	using the measure $\int{\cal D}\eta =\int \prod_{k=1}^{\kappa} \dd\eta_{k}$. The integral in Eq.~\eqref{avedyn2} simply gives the Lindblad decay dynamics $\varrho(\tau) = e^{{\cal L}\tau}\rho_0$. Therefore, we obtain the full solution for the average dynamics for this case, 
	\begin{equation}
	\rho_{\rm av}(t) = 
	\begin{cases}
	e^{ {\cal L} t}\rho_{0}, & t < \tau \\
	e^{{\cal K}(t-\tau)}\rho(\tau), & t \ge \tau
	\end{cases}, \label{case22}
	\end{equation}
	where ${\cal K}\bullet = -i [\hat{H}, \bullet]$ represents the generator for the pure unitary map. This solution shows that the feedback can partially interrupt the dephasing effect and restore the unitary dynamics; however, the quality of the restoration depends on the size of the delay time. One can see that the solution reduces to the pure unitary dynamics when $\tau \rightarrow 0$ and reduces to Eq.~\eqref{case11} for $\hat{H} = 0$.

	\begin{figure}[t]
\centering
    \subfloat[\label{fig1}]{ \includegraphics[width=0.48\linewidth]{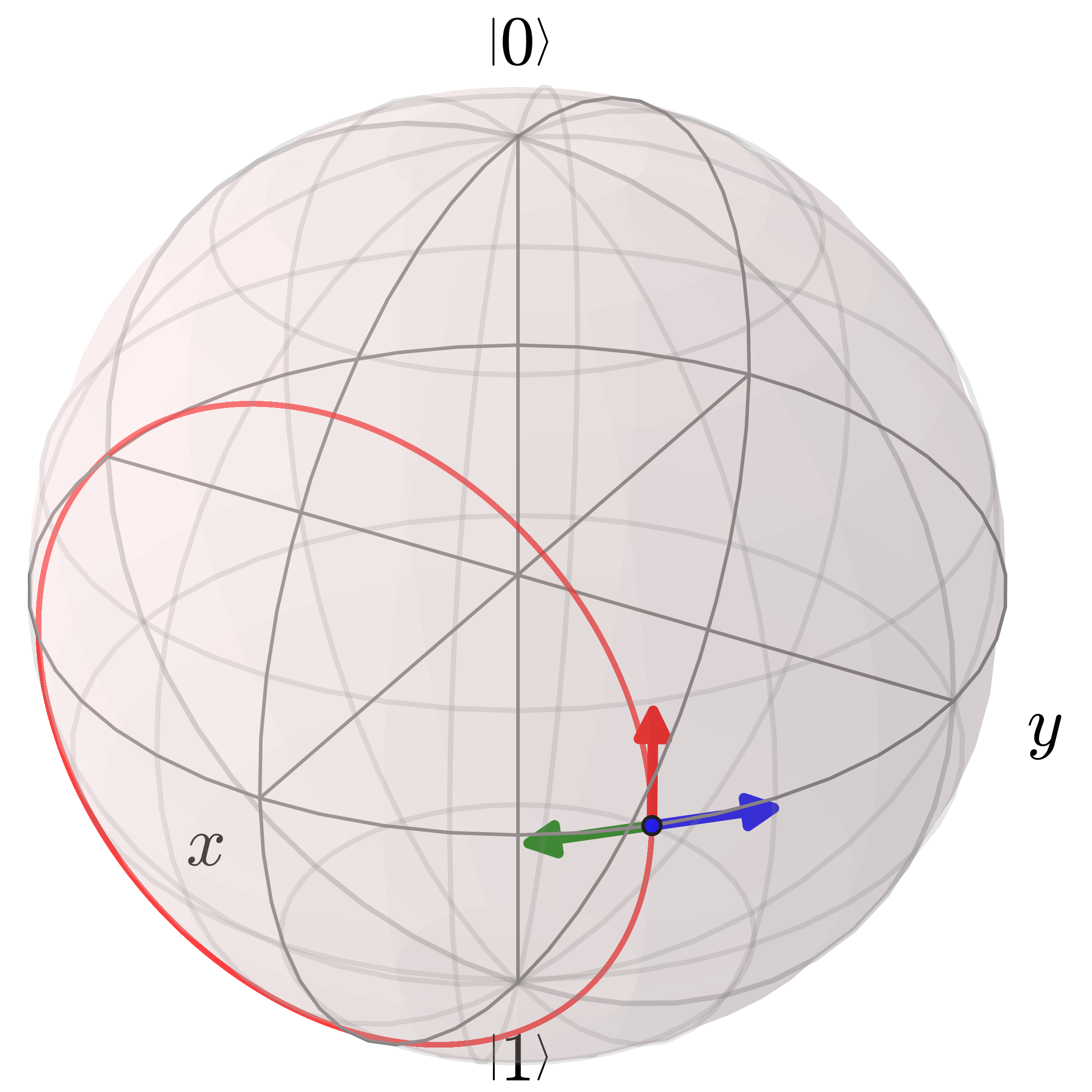} }
    \subfloat[\label{fig2}]{ \includegraphics[width=0.48\linewidth]{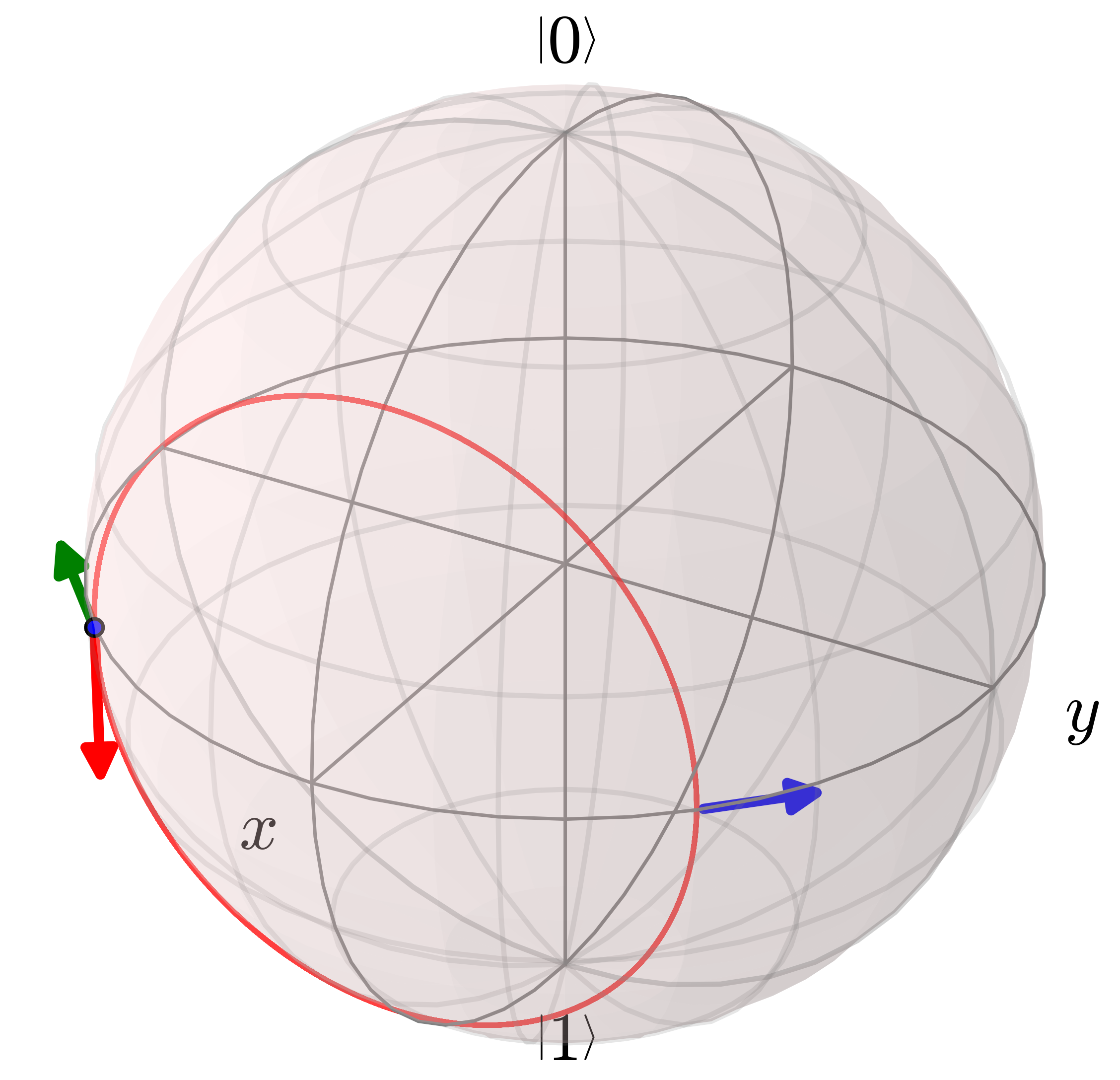} }
    \caption{Diagrams showing the two cases where the delayed feedback can suppress (panel (a)) or enhance (panel (b)) the decoherence. The unitary dynamics is presented with the red curves and arrows, while the dynamics from backaction and feedback are presented with blue arrows and green arrows, respectively. The decoherence suppression occurs when the feedback is delayed by an integer multiple of the Rabi period (a) and the decoherence enhancement occurs when the delay time is half integer multiple of the period.}
	\label{Fig:case34}
\end{figure}
	
	We show, in Fig.~\ref{Fig1:case2}, the qubit's average dynamics with the Rabi oscillation, $\hat{H} = \frac{\Omega}{2} \hat\sigma_z$, where we set $\Omega = 2\pi$. The effects of the delay time appear as the deviation of the average dynamics from the Lindblad decay (dashed red), occurring at the delay time $\tau = \alpha T_{\Omega}/2$, for $\alpha = 0.2$, $0.6$, and $1.0$. The revival of oscillation amplitudes are presented after that. The strength of the restored oscillation amplitude depends on the state $\rho(\tau)$ at time $\tau$, which depends on the size of the delay times. The oscillation can be restored close to the pure Rabi oscillation (high amplitude) when the delay time is small.

	\subsection{\label{subsec:3rd}Feedback process with non-commuting unitary dynamics}
	
	The last case we investigate is when the unitary dynamics does not commute with the measurement backaction, that is $[\hat{H}, \hat{L}] \ne 0$. Unlike the previous example, we cannot rearrange the operators from different time steps or combine them to cancel each other. Because of the non-commuting property, $[\hat{H}, \hat{L}] = \epsilon$, rearranging dynamical operators will create terms with polynomials of $\epsilon$ and terms of the order $\mathcal{O}(\ddt)$. Therefore, an analytic solution for this case is rather difficult to achieve. We then numerically investigate the qubit example, with a non-commuting Rabi oscillation given by $\hat{H} = \frac{\Omega}{2} \hat{\sigma}_{x}$, for the same coupling operator, $\hat{L} = \sqrt{\gamma}\hat{\sigma}_{z}$, as in the previous cases.

	Average dynamics of the qubit with the non-commuting unitary evolution is shown in Fig.~\ref{Fig1:case3}, using $5\times 10^3$ noise realizations, for different delay times $\tau = \alpha T_{\Omega}/2$, where $\alpha = 2.0$, $3.0$, $3.5$, $4.0$, and $5.0$. Here, the values of $\alpha$ and $T_{\Omega}$ are chosen such that the remarkably different effects on the decoherence dynamics can be easily seen.  The qubit's dynamics shows oscillation in the $y$-$z$ plane of the Bloch sphere, which is the same plane as the rotation dynamics caused by the measurement backaction. Differently from the previous case, the results show that the delayed feedback can either (a) partially restore he unitary dynamics (Rabi oscillation) or even (b) speed up the damping. As seen in Fig.~\ref{Fig1:case3}(a), the average dynamics in the $x$-coordinate can either be increased or reduced in size, depending on the delay time. For $\tau = T_\Omega$ and $2T_\Omega$ (green and brown circle dots in Fig.~\ref{Fig1:case3}), we can see the effect of restoration of the oscillation in Fig.~\ref{Fig1:case3}(b) as their amplitudes are slightly larger than that of the Lindblad evolution (dashed red). This similar pattern also happens for the delay times equal to other integer multiples of $T_\Omega$. However, for the delay times that are multiples of half the Rabi period, such as $\tau = 1.5 T_\Omega$ and $2.5 T_\Omega$, the oscillation amplitudes are smaller than that of the Lindblad decay, shown in pink and purple stars in Fig.~\ref{Fig1:case3}(b). Surprisingly, this damping effect could be even worse for shorter delay times (for example, see the purple stars in comparison to the pink stars).

	We can understand the two opposite effects on the oscillation amplitude from looking at the qubit's dynamics on the Bloch sphere (see Fig.~\ref{Fig:case34}). The Rabi oscillation makes the qubit's state rotate around the $x$-axis (shown as the red curves and red arrows), while the measurement backaction (blue arrows) and the feedback (green arrows) rotate the state around the $z$-axis, depending on the signs and values of the measurement results. Let us first consider an effect for a particular measurement result occurring when the state is at the blue dot in panel (a), where the measurement backaction creates a kick in the direction of the blue arrow. If there is no delay, the kick will be cancelled instantaneously with the feedback shown as the green arrow. If the delay time is an integer of the Rabi period $\tau = k T_\Omega$ for $k = 1, 2, ...$, the qubit state has evolved from the blue dot, but will still be roughly back to the initial point after the delay time (given that $k$ not too big). Therefore, the feedback kick can still move the state in the opposite direction and partially reverse the past effect of the backaction, though not as perfect as in the instantaneous case.  
	
	However, if the delay time is equal to half integers of the period, panel (b), the kick from the feedback operation (green arrow) will instead amplify the effect of the backaction and diverge the qubit path away from the correct Rabi path (moving towards states with lower values in the $x$-coordinate). This can be considered a similar effect as when an additional noise is added onto the system, making the qubit state decay further in the $x$-direction. Therefore, the delayed feedback for the no-knowledge measurement can either suppress or enhance the amplitude damping, depending on the system's dynamics and the delay time. We also note that the two opposite effects can occur in a more complicated setting, e.g., when a system's evolution has multiple time scales. As long as such a system has a well-defined periodic evolution, the incomplete decoherence suppression can still be achieved by setting the delay time equal to an integer of its evolution's period. 
	
	This result suggests that, in the case of slow feedback (the delay time is of the order of the oscillation period) or fast feedback is not possible, the experiments can be designed with specific finite delay times where the decoherence can still be partially suppressed. In our analyses, the delay times are defined in units of the dephasing time, $T_\gamma$, or the Rabi period, $T_\Omega$, so they can be conveniently converted to values in experiments. 
Recent examples with continuous measurements and feedback controls are the superconducting-qubit experiments~\cite{CamFlu2013,CamJez2016,NagTan2020}, where the detection chains and feedback operations typically add delay times of about 50-500 ns, for the dephasing time and the Rabi period in the order of micro-seconds. In that case, the delay time can be tuned to values as small as $\tau \approx 0.05 T_\gamma \sim 0.05 T_\Omega $.

	
	\section{\label{sec:Conslusion}Conclusion}
	We have presented the investigation on the time-delayed feedback with no-knowledge continuous measurement, and showed that finite delay times can degrade the decoherence suppression, even with perfect efficient detectors and commuting system's dynamics. We have performed theoretical analyses from the perspective of SMEs, modified with feedback delay times, and analytical solutions of the system's average dynamics. We showed results for three scenarios, with different combinations of measurement, feedback, and unitary evolution (both for commuting and non-commuting cases), and investigated numerical simulations for the driven qubit example. We have simulated the qubit trajectories and computed their average dynamics from an ensemble of stochastic realizations. 
	
	Since the measurement backaction causes the qubit to decohere on average, any imperfect backaction cancellation from the time-delayed feedback can result in an incomplete decoherence suppression. From our results of qubit's averaged trajectories, we find that the delay time can have significant effects on the decoherence suppression, i.e., degrading the stabilized states in the case of no unitary evolution (Fig.~\ref{Fig1:case1}) and degrading the Rabi oscillations in the case of commuting unitary dynamics (Fig.~\ref{Fig1:case2}). 
	
	Interestingly, when the qubit's unitary evolution (Rabi oscillations) do not commute with the measurement backaction, we find that the effect of delay time can either suppress or enhance the oscillation amplitude damping. We showed that the feedback needed to be turned on at the time equal to multiples of the Rabi period, in order to get the best partial decoherence cancellation. Otherwise, the decoherence effect could be worse, especially when the delay time is at half integers of the Rabi period, where the qubit's oscillation amplitude damped even faster than the Lindblad dynamics with no feedback control.

	For the final remarks, we emphasize that the decoherence suppression from this no-knowledge feedback is a result of measuring the environmental noise that affects the system and removing the noise backation on the system via the feedback control. The ability to measure noises that affect the system is not inherently quantum. It is also possible for classical stochastic systems. However, for the quantum cases, it is necessary that the measurement is ``no-knowledge," that is, the measurement result should not have any information about the measured system. Moreover, we note that there are also other factors in real experiments, such as measurement efficiencies and unmonitored noise channels, which can further degrade the decoherence suppression, and their effects are to be investigated further in future work.

	\begin{acknowledgments}
		We would also like to thank H. M. Wiseman for useful comments and discussions. J.S.~acknowledges the Faculty of Science, Mahidol University, for the Sri-Trang Tong scholarship for an opportunity to visit Centre for Quantum Dynamics, Griffith University, Australia. A.C.~acknowledges the support of the Griffith University Postdoctoral Fellowship scheme and the Australian Government via the AUSMURI grant AUSMURI000002. This research was supported by the Program Management Unit for Human Resources \& Institutional Development, Research and Innovation (Thailand) [grant number B05F630108] .
	\end{acknowledgments}
	
	\bibliographystyle{apsrev4-2}
	
%

	\appendix
	
	\section{Derivation}\label{sec:App}
	Here we derive the stratonovich and It\^o SMEs for the no-knowledge feedback with delay time. After the feedback is added, the quantum state is mapped by three operators: the measurement operator $\hat{M}_t = \exp[i\xi(t)\hat{L}\ddt]$, the unitary operator $\hat{U} = \exp[-i\hat{H}\ddt]$, and the feedback operator $\hat{F}_t = \exp[-i \xi(t-\tau) \hat{L}\ddt]$. The evolution of the normalized state is given by
	\begin{align}
	\rho(t+\ddt) = \frac{\hat{F}_{t}\hat{U}\hat{M}_{t}\rho(t)\hat{M}_{t}^{\dagger}\hat{U}^{\dagger}\hat{F}^{\dagger}_{t}}{{\rm Tr}\left[ \hat{F}_{t}\hat{U}\hat{M}_{t}\rho(t)\hat{M}_{t}^{\dagger}\hat{U}^{\dagger}\hat{F}^{\dagger}_{t}\right]}. \label{MAP}
	\end{align}
	The Stratonovich SME is obtained by expanding the above equation to the first order of an infinitesimal interval $\ddt$. The three operators become
	\begin{align}
	\hat{M} \approx & \, \mathrm{I} + i\xi(t)\hat{L}\ddt,\\
	 \hat{U} \approx  & \, \mathrm{I} - i\hat{H}\ddt, \\
	 \hat{F}_{t} \approx &\,  \mathrm{I} - i\xi(t-\tau)\hat{L}\ddt,
	\end{align}
	and the combination of all three is given by
	\begin{align}
	\hat{F}_{t}\hat{U}\hat{M}_{t} = \mathrm{I} - i\hat{H}\ddt + i[\xi(t) - \xi(t-\tau)]\hat{L}\ddt + \mathcal{O}(\ddt^2).
	\end{align}
	Putting these approximations into Eq.$\eqref{MAP}$, keeping terms up to $\mathcal{O}(\ddt)$, and taking the continuum limit $\ddt \rightarrow \dt$, we get
	\begin{align}
	\dot{\rho}(t) =& -i[\hat{H} - [\xi(t)-\xi(t-\tau)]\hat{L}, \rho(t)],
	\end{align}
	as the Stratonovich SME for a finite delay time $\tau$.
	
	On the other hand, for the It\^o SME, we need to consider the It\^o rule, which comes from the Wiener's process defined by $\dd W(t) = \xi(t)\dt$. Therefore, when expanding the operators in Eq.~\eqref{MAP}, we need to keep in mind that the Wiener process has an important property, $(\dd W(t))^{2} = (\xi(t)\ddt)^{2} \approx \ddt$ in the mean-square limit. Therefore, the measurement and feedback operators become
	\begin{align}
	\hat{M}_{t} \approx &\,   \mathrm{I} + i\xi(t)\hat{L}\ddt - \frac{1}{2}\hat{L}\hat{L}\ddt,\\
	 \hat{F}_{t} \approx & \, \mathrm{I} - i\xi(t-\tau)\hat{L}\ddt - \frac{1}{2}\hat{L}\hat{L}\ddt.
	\end{align}
	As a result, substituting these back in Eq.~\eqref{MAP} and applying It\^o rule whenever possible, this gives the It\^o SME for the delayed feedback,
	\begin{equation}
	\begin{aligned}
	\rho(t+\ddt) = {} &\{\mathrm{I} - i\hat{H}\ddt + i[\xi (t) - \xi (t-\tau)]\hat{L}\ddt - \frac{1}{2}\hat{L}\hat{L}\ddt \} \\ & \rho(t) \{\mathrm{I} + i\hat{H}\ddt - i[\xi (t) - \xi (t-\tau)]\hat{L}\ddt - \frac{1}{2}\hat{L}\hat{L}\ddt \},
	\end{aligned}
	\end{equation}
	which becomes Eq.~\eqref{E2} in the main text
	\begin{align}
	d\rho(t) =&  -i[\hat{H} - [\xi(t)-\xi(t-\tau)]\hat{L}, \rho(t)]\ddt\\
	& - 2\mathcal{D}[L]\rho(t)[\xi (t) \xi (t-\tau) \ddt^{2} - \ddt],
	\end{align}
	in the time-continuum limit $\ddt \rightarrow \dt$.

\end{document}